\documentclass[%
 reprint,
 amsmath,amssymb,
prl,
 longbibliography,
 lengthcheck,%
]{revtex4-1}

\usepackage{graphicx}% Include figure files
\usepackage{dcolumn}% Align table columns on decimal point
\usepackage{bm}% bold math
\usepackage{hyperref}% add hypertext capabilities
\usepackage{txfonts}
\makeatletter

\newcommand{\Rmnum}[1]{\expandafter\@slowromancap\romannumeral #1@}
\makeatother

\begin{document}

\preprint{APS/123-QED}

%SrFeAsF
\title{Charge redistribution at the antiferromagnetic phase transition in SrFeAsF compound}% Force line breaks with \\

%\author{Zhiwei Li}
%\author{Xiaoming Ma}
%\author{Xin Liu}
%\author{Tao Wang}\email{lizhiwei03@lzu.cn}
\author{Zhiwei Li}%\email{lizhiwei03@lzu.cn}
\author{Yang Fang}
\author{Xiaoming Ma}
\author{Hua Pang}\email{hpang@lzu.edu.cn}
\author{Fashen Li}%\email{lifs@lzu.edu.cn}
 \affiliation{Institute of Applied Magnetics, Key Lab for Magnetism and Magnetic Materials of the Ministry of Education, Lanzhou University, Lanzhou 730000, Gansu, P.R. China.}%Lines break automatically or can be forced with \\

%\author{Xianhui Chen}
% \affiliation{Hefei National Laboratory for Physical Science at Microscale and Department of Physics, University of Science and Technology of China, Hefei 230026, Anhui, P.R. China. }

\date{\today}% It is always \today, today,

\begin{abstract}
The relationship between spin, electron, and crystal structure has been one of the foremost issues in understanding the superconducting mechanism since the discovery of iron-based high temperature superconductors. Here, we report M\"ossbauer and first-principles calculations studies of the parent compound SrFeAsF with the largest temperature gap ($\sim$50\,K) between the structural and antiferromagnetic (AFM) transitions. Our results reveal that the structural transition has little effect on the electronic structure of the compound SrFeAsF while the development of the AFM order induces a redistribution of the charges near the Fermi level.

\begin{description}
  %  \item[Usage]
  % Secondary publications and information retrieval purposes.
\item[PACS numbers]
76.80.+y, 74.10.+v, 71.15.Mb
  %  \item[Structure]
  %  You may use the \texttt{description} environment to structure your abstract;
  %  use the optional argument of the \verb+\item+ command to give the category of each item.
\end{description}
\end{abstract}

\maketitle

%\tableofcontents

\section{\label{sec:Intro}Introduction}

The discovery of superconductivity (SC) in iron-based superconductors, like the cuprates, promises to be an important milestone in condensed matter physics \cite{JACS-Hosono,Review-Greene,Review-HHWen}. Proximity to a magnetically ordered state has been suggested to be a crucial factor for the observed high-temperature (high-$T_C$) SC \cite{Review-Magnetism}. Another interesting feature is that the antiferromagnetic (AFM) transition is always accompanied by a structural transition either before the AFM transition or at the same temperature (see \cite{ThePuzzle} and references therein). In the 122-type AeFe$_2$As$_2$ (Ae=Ba, Sr, Ca, Eu) parent compounds and the 11-type Fe$_{1+\delta}$Te systems, the structural and AFM transitions occur simultaneously. For the rare-earth based 1111-type ReFeAsO (Re=La, Ce, Pr, Nd, etc.) materials and the 111-type AFeAs (A=Li, Na) compounds, the structural transition precedes the AFM transition by about 10$\sim$20\,K. Till now, the relationship between the structural and magnetic phase transition has been an open question though it has been suggested that the structural transition may be driven magnetically \cite{MdriveS}. But there is angle-resolved photoemission spectroscopy (ARPES) evidence \cite{PRL-Relation} that structural and AFM transitions may share the same origin and could both be driven by the electronic structure reconstruction in NaFeAs. Actually, the nature of the AFM order itself is still under debate, with some favoring a localized picture and others supporting an itinerant nature.

To understand the interplay between structural, magnetic, and electronic properties, a microscopic investigation by M\"ossbauer spectroscopy \cite{SDW-shape,CompeteMSC,1111-Moss,Moss-1111Gd,ZhiweiLi,Moss-Review} may be very helpful. In the present work, we report detailed M\"ossbauer spectroscopy measurements and first-principles calculations of SrFeAsF. The SrFeAsF parent compound, which has the largest reported temperature interval ($\sim$50\,K) between the two phase transitions \cite{TS-175K}, offers an ideal opportunity to address what happens between the structural and AFM transitions. Our results show strong evidence of charge redistribution at the AFM phase transition in the SrFeAsF compound.

\section{\label{sec:Experiment}Experiments}
SrFeAsF was synthesized using a two-step solid-state reaction method, which is similar to that described earlier \cite{Preparation-PRB}. First, SrAs/FeAs was prepared by reacting Sr/Fe flakes/powders and As powders sealed in an evacuated quartz tube at 773\,K for 10\,h and then at 1000\,K for 20\,h. The resultant precursors, SrAs, FeAs, SrF$_2$ and Fe powders, were mixed and thoroughly ground at a molar ratio of 1:1:1:1. All the weighing and mixing procedures were performed in a glove box with protective argon atmosphere. Then the mixture was pressed into a pellet, sealed in an evacuated quartz tube and heated up to 1273\,K at a rate of 60\,K/h, kept at this temperature for 40\,h, and cooled down to room temperature. The product was homogenized in an agate mortar, pressed into a pellet, and sintered at 1273\,K for 40\,h again to obtain the final sample.

Phase purity was checked by x-ray powder diffraction (XRPD) on a Philips X'pert diffractometer with Cu K$_\alpha$ radiation. Rietveld refinement of the SrFeAsF was performed with the GSAS package \cite{GSASpackage} to obtain the lattice parameters. Transmission M\"ossbauer spectra (MS) at temperatures between 16 and 290\,K were recorded using a conventional constant acceleration spectrometer with a $\gamma$-ray source of 25\,mCi $^{57}$Co in a palladium matrix moving at room temperature. The absorber was kept static in a temperature-controllable cryostat filled with helium gas. The drive velocity is calibrated with sodium nitroprusside (SNP) at room temperature and all the isomer shifts quoted in this work are relative to that of the $\alpha$-Fe.

\section{\label{sec:Calculations}Calculation Methods}

We use the full potential linearized augmented plane wave (FP-LAPW) \cite{FP-LAPW} method as embodied in the WIEN2K \cite{WIEN2k,WIEN2kWeb} code in a scalar relativistic version without spin-orbit coupling. According to the FP-LAPW method, the simulation cell is divided into spherical atomic regions with radii $\mathbf{R}_{mt}$ (2.30, 2.41, 2.13, and 2.30 a.u. for Sr, Fe, As, and F atoms, respectively.) and an interstitial region. The wave functions are described by atomic like functions inside the atomic spheres, while in the interstitial region plane waves are used. The maximum angular momenta $l$ for the expansion of the wave functions in the spherical harmonics inside the spheres is confined to $l_{max} = 10$. The cutoff parameter $\mathbf{R}_{MT}\mathbf{K}_{MAX}$ for limiting the number of the plane waves is set as 7.0. The magnitude of the largest vector in the Fourier expansion of the electron density ($\mathbf{G}_{max}$) is set to 14.0. States lying more than 6Ry below the Fermi level are treated as the core states. The Brillouin zone integration is done with a modified tetrahedron method using 648 special $k$-points in the irreducible wedge. The generalized gradient approximation (GGA) suggested by Perdew, Burke and Ernzerhof (PBE GGA) \cite{GGA-PBE} is employed for the exchange-correlation effects.

Once the electron (spin-) densities are calculated self-consistently with high accuracy, hyperfine parameters can be deduced directly \cite{Calc-EFG,Calc-Moss}. The electric field gradient (EFG) tensor can be obtained from an integral over the nonspherical electron density. The most interesting principal component $V_{zz}$ is given by
$V_{zz}=\frac{1}{2\pi\varepsilon_0}\int_0^{2\pi}d\phi\int_0^{\pi}d\theta\sin\theta P_2(\cos\theta)\int_0^{\infty}\frac{dr}{r}\rho(\textbf{r}),$
where $P_2$ is the second-order Legendre polynomial and $\rho(\textbf{r})$ is the electron density. In the LAPW method, $\rho(\textbf{r})$ is expressed by lattice harmonics, i.e. under an $LM$ form, $\rho_{LM}(\textbf{r})=\sum\limits_{E_{nk}<E_F}\sum\limits_{l,m}\sum\limits_{l',m'}R_{lm}(\textbf{r})R_{l'm'}(\textbf{r})G_{Lll'}^{Mmm'}$. It turns out, only the component $\rho_{20}$ is important to $V_{zz}$. And when $L=2$ and $M=0$, only $l=l'=1$ and $l=l'=2$ (and to a minor extent $l=0$, $l'=2$ and $l=1$, $l'=3$) give non-zero Gaunt numbers, which correspond to $p$-$p$ and $d$-$d$ contributions to the EFG, respectively.

\section{\label{sec:Results}Results and Discussion}
\begin{figure}[htp]
\includegraphics[width=8 cm]{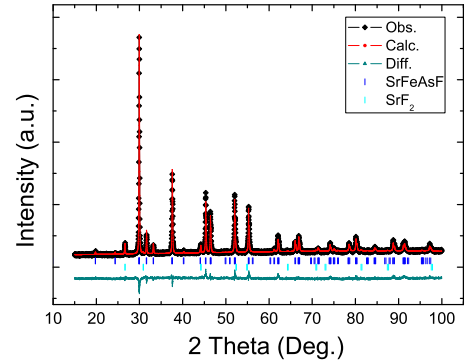}% Here is how to import EPS art
\caption{\label{XRPD}(Color online) X-ray powder diffraction pattern of SrFeAsF ($\Diamondblack$) and Rietveld fit (red solid line). Bars at bottom show the calculated Bragg diffraction positions of SrFeAsF (upper row) and a small amount of SrF$_2$ (lower row) impurity phase (less than 4\%). The olive green line is the difference between the observed and calculated patterns.}
\end{figure}

Figure \ref{XRPD} shows the XRPD pattern of the sample, from which one can see that all the main peaks can be indexed to the SrFeAsF phase with a ZrCuSiAs-type structure. The lattice constants were determined to be a=4.002\,{\AA} and c=8.970\,{\AA}, which is in good agreement with previously reported values \cite{Preparation-PRB}. Only a small amount of the SrF$_2$ impurity phase was detected, which is determined by the Rietveld refinement to be less than 4\%. As the impurity phase (SrF$_2$) does not contain iron, the analyses of the MS can not be affected and this is shown in what follows.

\begin{figure}[htp]
\includegraphics[width=8 cm]{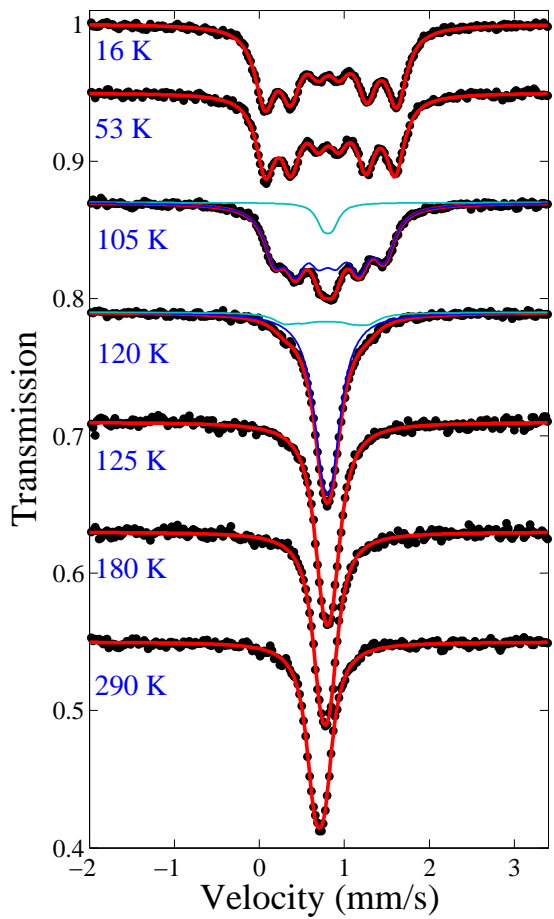}% Here is how to import EPS art
\caption{\label{Moss}(Color online) M\"ossbauer spectra taken at indicated temperatures of the SrFeAsF sample together with transmission integral fits (solid line).}
\end{figure}

M\"ossbauer spectra recorded at indicated temperatures together with transmission integral fits are shown in Fig. \ref{Moss}. Clearly, the spectra change from the paramagnetic (PM) state (singlet or unresolved doublet) to the magnetically split state (sextet) upon lowering of the temperature. The Hamiltonian for the hyperfine coupling in the principal axis system of the EFG can be expressed as \cite{MossBookChen}
{%\footnotesize
\begin{eqnarray}
\mathcal{\hat{H}}_{QM}&& = \mathcal{\hat{H}}_Q + \mathcal{\hat{H}}_M \\ \nonumber
&& = \frac{eQV_{zz}}{4I(2I-1)}[3\hat{I}^2_z - \hat{I}^2 + \eta(\hat{I}_x^2 - \hat{I}_y^2)] \\ \nonumber
&& - g\mu_NB[(\hat{I}_x\cos\phi + \hat{I}_y\sin\phi)\sin\theta + \hat{I}_z\cos\theta],
\end{eqnarray}
}where $\hat{I}$ and $\hat{I}_x, \hat{I}_y, \hat{I}_z$ refer to the nuclear spin operator and operators of the nuclear spin projections onto the principal axes and $Q$ denotes the quadrupole moment of the nucleus (for excited $^{57}$Fe $Q = + 0.17(1)$\,b \cite{Q-PRB76155118}). $\phi$ and $\theta$ are the azimuthal and polar angles of the hyperfine magnetic field in the EFG coordinate system, respectively. $\eta=(V_{xx}-V_{yy})/V_{zz}$ represents the asymmetry parameter of the EFG at the nucleus. In the present case, it is reasonable to assume that the EFG has axial symmetry around the $c$-axis \cite{1111-Moss}. Therefore, above the AFM transition temperature, one has a doublet with quadrupole splitting $\Delta E_Q = \frac{1}{2}|eQV_{zz}|$.
Below the AFM transition temperature, due to the smallness of the quadrupole interaction in comparison with the magnetic dipole interaction, the assumption of first-order perturbation can be made. Hence, eigenvalues of the magnetic Hamiltonian are perturbed (shifted) by the term $\varepsilon = (-1)^{|m_I|+1/2}(eQV_{zz}/8)\cdot(3\cos^2\theta - 1)$ \cite{MossBookChen}. Considering the established magnetic structure, $\theta=90^{\circ}$, the sign of $V_{zz}$ and the nuclear quadrupole coupling constant (NQCC), $|eQV_{zz}|$, can be obtained subsequently.

As can be seen, the spectra can be well fitted by only one doublet/(sextet) above/(below) the transition region, indicating that the local environment of the Fe ion is unique. This means that there is no Fe containing impurity phase in our sample, coincident with the XRPD results. At room temperature the best fit of the spectra gives an isomer shift of $\delta =$0.44\,mm/s and a quadrupole splitting of $|\Delta E_Q|=$0.12\,mm/s, which compares well with reported values for SrFeAsF \cite{SrFeAsF-Moss} and other 1111-type iron-pnictides \cite{1111-Moss,Moss-1111Gd}. At temperatures in the transition region (100\,K-125\,K) the MS are fitted with the superposition of a doublet and a sextet \cite{1111-Moss,Moss-1111Gd} as usually done in the M\"ossbauer studies of the iron-based parent compounds.

\begin{figure}[htp]
\includegraphics[width=8 cm]{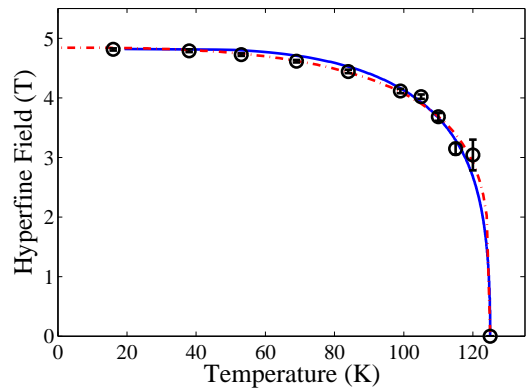}% Here is how to import EPS art
\caption{\label{Bhf}(Color online) Temperature dependence of the hyperfine field, $B_{hf}(T)$, extracted from transmission integral fits of the M\"ossbauer spectra. The red dashdotted line and the blue solid line are fits to the data with the power law and the Bean-Rodbell model, respectively (see text).}
\end{figure}

The temperature dependence of the hyperfine field, $B_{hf}(T)$, is shown in Fig. \ref{Bhf}. Fitting $B_{hf}(T)$ data to the usual function $B_{hf}(T)=B_{hf}(0)[1-(T/T_N)^{\alpha}]^{\beta}$ \cite{TS-175K} (red dashdotted line) leads to $B_{hf}$(0)=4.84\,T, $T_N$=125\,K, $\alpha=2.8$ and $\beta=0.23$. These values agree very well with those derived from fitting the upper precession frequency of the $\mu$SR data \cite{TS-175K} and prove the two-dimensional magnetism \cite{TwoD-beta} in the SrFeAsF compound. To find out the order of the magnetic phase transition, we fit the $B_{hf}(T)$ data according to the Bean-Rodbell model (blue solid line) \cite{ZhiweiLi,Bean-Rodbell} for $J=1/2$, $T/T_0=(\sigma/\tanh^{-1}\sigma)(1+\zeta\sigma^2/3)$, where $T_0$ is the transition temperature if the lattice is not compressible, $\sigma$ is the sublattice magnetization, and $\zeta$ is a fitting parameter ($\zeta<1$ for a second order transition, $\zeta>1$ for a first order transition and for $\zeta=0$ the equation reduces to a Brillouin function). The fitted values for $\zeta$, $T_0$ and $B_{hf}(T)$ are found to be 0.8, 125\,K and 4.82\,T, respectively. Obviously, $\zeta$ is a little smaller than 1, which is different from that obtained for CaFe$_2$As$_2$ ($\zeta=1.35$), indicating that the AFM transition for SrFeAsF is more second-order-like than that for CaFe$_2$As$_2$, which is in agreement with previous reports \cite{TS-175K,SrFeAsF-Moss}.

\begin{figure}[htp]
\includegraphics[width=8 cm]{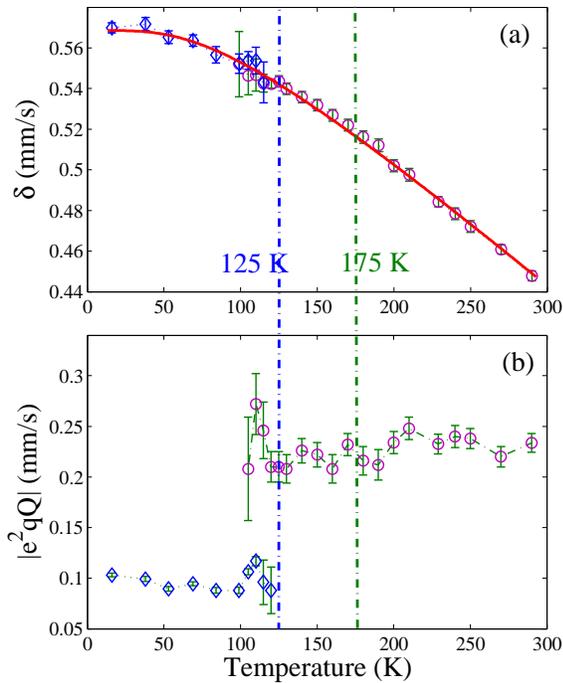}% Here is how to import EPS art
\caption{\label{QSIS}(Color online) Temperature dependence of the isomer shift $\delta$ together with theoretical fits using the Debye model, (a), and the nuclear quadrupole coupling constant, (b), of the SrFeAsF compound. The blue-diamonds (purple-circles) are derived from the AFM (PM) phase. Vertical dash-dotted lines at 125 and 175\,K \cite{TS-175K} indicate the AFM and structural transition temperatures, respectively.}
\end{figure}

Temperature dependence of the isomer shift, $\delta(T)$, determined from the fits of the MS is shown in Fig. \ref{QSIS} (a). The two vertical dashdotted lines at 125 and 175\,K \cite{TS-175K} indicate the AFM and structural transition temperatures, respectively. As expected, $\delta(T)$ increases gradually with decreasing temperature due to the second order Doppler effect. And in terms of the Debye model, the Debye temperature is found to be $\Theta_D$=337\,K, which is in reasonable agreement with those determined from specific-heat measurement, $\Theta_D$=339\,K \cite{SrFeAsF-Moss}, and neutron diffraction experiment, $\Theta_D$=347\,K \cite{SrFeAsF-neutron}. In Fig. \ref{QSIS} (b) we present the temperature dependence of the NQCC, $|eQV_{zz}|$, which is obtained as $2\Delta E_Q$ and $8|\varepsilon|$ for PM and AFM states, respectively.

Obviously, no abrupt changes of $\delta(T)$ and NQCC during the structural transition around $\sim$175\,K are found, implying a rather continuous structural transition. In terms of group theory, the space group of the low-temperature phase (Cmma) is a subgroup of that of the high-temperature phase (P4/nmm) and therefore a relatively continuous structural transition is usually the case \cite{SrFeAsF-Moss}. Considering the small change in volume after the transition \cite{SrFeAsF-neutron,SrFeAsF-Moss}, it is likely that the charge density at the iron nucleus and the spatial distribution symmetry of the charges exhibit negligible changes. Therefore, one may not expect distinct changes in $\delta(T)$ and NQCC near the structural transition temperature range. Interestingly, upon cooling through the AFM transition, the NQCC jumps to a much smaller value for the AFM phase while the temperature dependence of $\delta(T)$ still follows the Debye model without any evident jump. The isomer shift changes rather smoothly with decreasing temperature through the AFM transition was also observed in the GdFeAsO compound \cite{Moss-1111Gd} in contrast to that for the 122-type CaFe$_2$As$_2$ material where a small jump was detected \cite{ZhiweiLi}. This could be understood as due to some charge redistribution effect, which changes the EFG at the iron nucleus while has little effect on the absolute charge density at the iron nucleus. In fact, charge redistribution effect have also been observed in 122-type AeFe$_2$As$_2$ \cite{122-recon} and 111-type NaFeAs \cite{PRL-Relation} systems and high-T$_C$ cuprates materials \cite{Cu-ReconPrl,Cu-ReconPrb}.

To get a better understanding of the charge redistribution effect, we calculate the EFG and the electronic properties of the SrFeAsF compound. For tetragonal (P4/nmm) SrFeAsF, the experimental lattice parameters and internal atomic positions (a=b=3.9930\,\AA, c=8.9546\,\AA, Z$_{Sr}$=0.1583, Z$_{AS}$=0.3485) are adopted \cite{TS-175K}. For orthorhombic (Cmma) SrFeAsF, the experimental value (a=5.6155\,\AA, b=5.6602\,\AA, c=8.9173\,\AA, Z$_{Sr}$=0.1584, Z$_{As}$=0.3475) are employed \cite{TS-175K}. The AFM phase of SrFeAsF is also set up according to experimental reports, that is, the spin direction is opposite between two adjacent FeAs layers and opposite along the longer Fe-Fe direction within the FeAs layers \cite{SrFeAsF-neutron}.

First, we have calculated the EFG principal component $V_{zz}$ for nonmagnetic-P4/nmm (NM-P4/nmm), NM-Cmma and AFM-Cmma phases of the SrFeAsF parent compound. The experimental and numerical results are shown in Table I. The sign of $V_{zz}$ for NM-P4/nmm and NM-Cmma phases can not be determined experimentally and the absolute values are given. For the magnetically split AFM state, the sign of $V_{zz}$ can be established directly in this case to be negative, which agrees well with first-principles calculations. This is different with other iron-based superconductors, where the sign of $V_{zz}$ is obtained to be positive \cite{1111-Moss,Moss-1111Gd,ZhiweiLi}. The reason for this difference certainly deserves future investigations. As can be seen, if we consider the absolute values of $V_{zz}$, the tendency of the calculated results match reasonably well with the experimental results. The $|V_{zz}|$ values are almost the same for NM-P4/nmm phase and the NM-Cmma phase. While the $|V_{zz}|$ value for the AFM-Cmma phase is significantly smaller than the value for the other two phases. Generally, the EFG is sensitive to structural changes and is taken as a good probe for structural transitions. Our observations, however, indicate that the structural transition from the P4/nmm phase to the Cmma phase in SrFeAsF has no obvious effect on the EFG, but the introduction of long-range magnetic order unambiguously reduces the $|V_{zz}|$ value of SrFeAsF.

\begin{table}[ht]
\centering
\caption{Principal component of the EFG, $V_{zz}$, for NM-P4/nmm, NM-Cmma and AFM-Cmma phases of the SrFeAsF compound. The experimental values for NM-P4/nmm and NM-Cmma phases are given as $|V_{zz}|$ (see text).}
\label{TableVzz} {
\begin{tabular}{c c c c}
\hline \hline
$V_{zz}$ (10$^{21}$\,V/m$^2$)  &    NM-P4/nmm   &  NM-Cmma         &   AFM-Cmma     \\
\hline
Calculation     &  -0.350       &  -0.330          &  -0.197         \\
Experimentation &  0.65(3)       &  0.64(4)          &  -0.27(1)        \\
\hline \hline
\end{tabular}}
\end{table}

\begin{figure}[htp]
\includegraphics[width=8 cm]{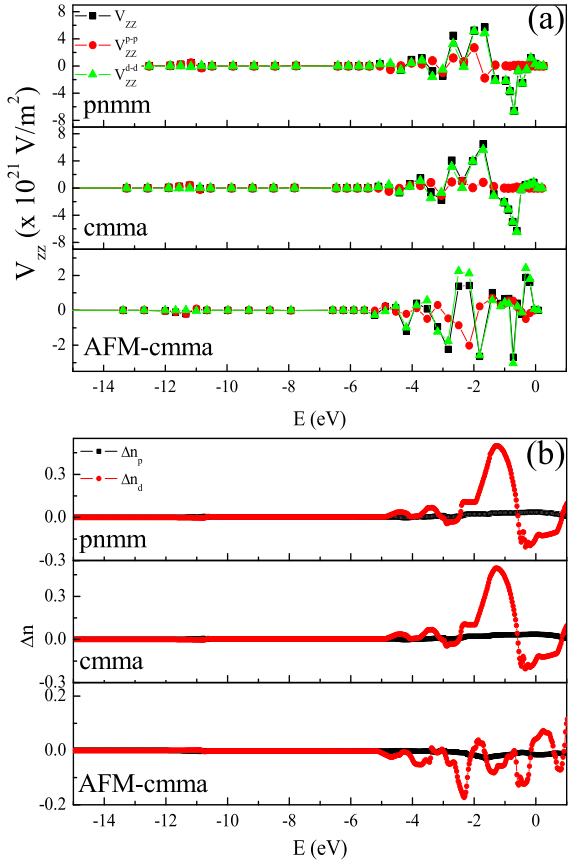}% Here is how to import EPS art
\caption{\label{Calc1}(Color online) Evolutions of $V_{zz}^{p-p}$, $V_{zz}^{d-d}$, and $V_{zz}$ (a) and, evolutions of $\Delta n_p$ and $\Delta n_d$ (b) with the state of energy (see text).}
\end{figure}

In order to further understand the $V_{zz}$, the evolutions of $V_{zz}^{p-p}$, $V_{zz}^{d-d}$ and $V_{zz}$ with the state of energy are investigated minutely and the results are shown in Fig. \ref{Calc1} (a). For all three structures, $V_{zz}^{p-p}$, $V_{zz}^{d-d}$ and $V_{zz}$ are nearly zero until around -5\,eV starting from the lower energy side, afterwhich $V_{zz}^{p-p}$ and $V_{zz}^{d-d}$ vibrate strongly until the Fermi energy is reached. Obviously, the $V_{zz}$ of all systems are dominated by the valence electrons near the Fermi surface. And the similarity in the evolution of $V_{zz}$ with energy is quite distinct for the two NM systems, which explains the experimental observation quite well. It is found that, though both $p$-$p$ interaction and $d$-$d$ interaction contribute to the $V_{zz}$, the contribution from $V_{zz}^{d-d}$ dominates.

\begin{figure}[htp]
\includegraphics[width=8 cm]{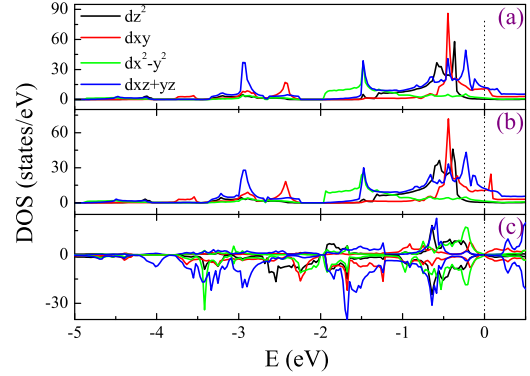}% Here is how to import EPS art
\caption{\label{Calc2}(Color online) Density of states of the iron atom. (a) NM-P4/nmm, (b) NM-Cmma and (c) AFM-Cmma as described in the text.}
\end{figure}

According to the above definition, the EFG is decided by the non-spherical electron-density $\rho_{20}(\textbf{r})$, as well as by the radial dependence $1/r^3$ of the anisotropic charge density. We calculated the nonspherical numbers of $p$ and $d$ charges inside the $\mathbf{R}_{MT}$ of the Fe atom, denoted as $\Delta n_p$ and $\Delta n_d$, respectively. $\Delta n_p$ and $\Delta n_d$ quantify the deviations of the charge distribution from spherical symmetry and are calculated as: $\Delta n_p = (n_{p_x}+n_{p_y})/2-n_{p_z}$ and $\Delta n_d = (n_{d_{xy}}+n_{d_{x^2-y^2}})-(n_{d_{xz}}+n_{d_{yz}})/2-n_{d_{z^2}}$ \cite{NonSphNumb}. The evolutions of $\Delta n_p$ and $\Delta n_d$ with the state of energy are shown in Fig. \ref{Calc1} (b). Obviously, the fluctuation of $\Delta n_d$ is much more pronounced than that of $\Delta n_p$, indicating stronger anisotropic spatial distribution of Fe-3$d$ electrons compared with that of the Fe-3$p$ electrons. This can be explained by the strong hybridization between Fe-3$p$ and As-4$s$ orbits, which reduces the spatial anisotropy of the Fe-3$p$ charges. The decrease of $\Delta n_d$ in the AFM phase can be understood from the calculated density of states (DOS) of the iron atom for the three structures. As shown in Fig. \ref{Calc2}, the partial DOS of Fe-$3d$ sub-orbitals for the two NM phases almost has the same graphics features in the energy range from -5.0\,eV to the Fermi level. Though there exists some differences above the Fermi level, the discrepancies do not affect the EFG. The differences in the DOS of the AFM state are quite distinct. Obviously, the $d_{xy}$, $d_{xz}$, and $d_{yz}$ orbitals control the symmetry of the Fermi surface in NM phases, yet the DOS is much reduced in the AFM phase, resulting in a much smaller Fermi surface \cite{JETP-calc}. Besides, the gravity centers of all five $d$ orbitals are closer to each other in energy, yielding more isotropic spatial distribution of $d$ electrons. This will reduce the anisotropy of charge density of $d$ electrons, hence resulting in a smaller EFG, proving the above observation.

The fact that $V_{zz}$ of the three systems is dominated by $d$-$d$ interactions is the key point to understand the sudden drop in the EFG of the MS observations. It is well known that the magnetic moment of the AFM-Cmma phase mainly comes from the spin polarization of Fe-$3d$ electrons. Due to the exchange splitting of the $3d$-band, the chemical potential of $3d$-electrons would decrease in a magnetic phase. Consequently, there should be a redistribution of electrons between $sp$- and $d$-bands. The charge redistribution due to spin polarization of Fe-$3d$ electrons may have at least two effects on the electronic structure of the AFM phase. First, the charge will transfer from Sr-F layers to Fe-As layers. Second, the $3d$ electrons will redistribute among the $3d$ suborbitals.

\begin{figure}[htp]
\includegraphics[width=8 cm]{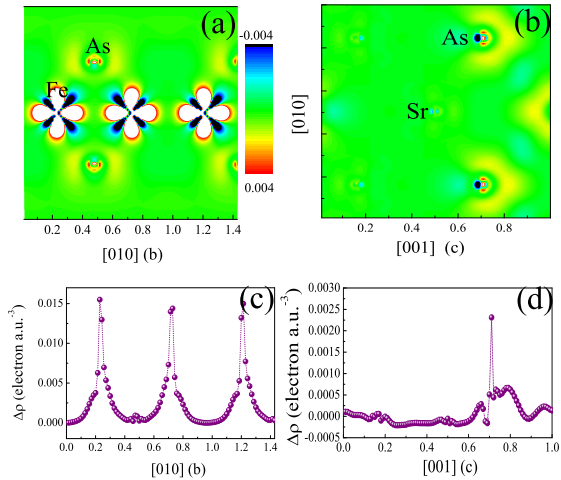}% Here is how to import EPS art
\caption{\label{Calc3}(Color online) The difference in electron density distribution between the NM-Cmma phase and the AFM-Cmma phase of the SrFeAsF compound in the Fe-As plane (a) and the Sr-As plane (b) and the profiles of the density difference along the [010] direction (c) and the [001] direction (d) of the crystal.}
\end{figure}

The calculated electron density differences of the NM-Cmma phase and the AFM-Cmma phase prove our suggestions. Figure \ref{Calc3} (a) displays the electron density difference $\Delta \rho$ for the two phases in the plane defined by the Fe atom and the two nearest As neighbors. The direction of the nearest Fe atoms is along the crystalline $b$-axis. The profile of the density difference along the $b$-axis is also shown in Fig. \ref{Calc3} (c), where a positive number denotes increased density after magnetization and a negative number denotes the decreased density. As can be seen, the most distinct increase in charge density happens at the iron sites, indicating the electron charges tend to accumulate at the iron sites after the emergence of magnetic ordering. We also calculated the electron density difference of the $yz$-plane with As and Sr atoms. As shown in Figs. \ref{Calc3} (b) and (d), the electrons distributed in the region between two adjacent Fe-As layers decrease after magnetic ordering, which means the electrons tend to flow from Sr-F layers to Fe-As layers accompanying the emergence of magnetic ordering.

Moreover, it is found that the density difference distribution around the iron site clearly has spatial characters of $3d$ orbitals, suggesting the redistribution of the electrons among the $3d$ orbitals. From the calculated occupation number of each orbital, it is found that electrons move from $d_{z^2}$, $d_{xz}$ and $d_{yz}$ orbitals to $d_{xy}$ and $d_{x^2-y^2}$ orbitals when the system goes into the magnetic state. Indeed, the electron density projected on the Fe-As plane increases along the $b$-axis after magnetization, as shown in Fig. \ref{Calc3} (a). While the electron density along the direction of the As-Fe bond decreases. This observation highly suggests that magnetization will enhance the localization of the $3d$ electrons, hence the correlation among $3d$ electrons.

Actually, magnetization induced charge redistribution is common in terms of thermodynamics. But it is not easy to observe experimentally in iron based superconductors \cite{PRL-Relation,122-recon} due to the interference of the structural phase transition, which happens close to the magnetic phase transition and may have magnetic origin \cite{MdriveS}. Luckily, in the SrFeAsF compound, the large temperature interval between the structural and magnetic phase transitions ($\sim$50\,K) makes it possible to exclude the disturbance in electronic redistribution due to the structural phase transition. For the SrFeAsF compound, Fe-$3d$ electrons contribute to the EFG as well as to the magnetism. And the EFG is sensitive to the spatial distribution of charges near the iron nucleus; accordingly, the charge redistribution around the iron site, due to the appearance of long-range magnetic order, can be detected by the EFG. These findings may also hold for other iron-pnictides and may shed light on understanding the fascinating physics of these materials.

\section{\label{sec:Conclusion}Concluding remarks}
In summary, we have studied the SrFeAsF parent compound by M\"ossbauer spectroscopy and first-principles calculations. Our results about the structural and AFM transitions are in good agreement with previous reports. The temperature dependence of the isomer shift has no anomalies during the two phase transitions, indicating no abrupt changes of the absolute charge density at the iron nucleus. Additionally, the structural transition has little effect on the main component of the EFG, $V_{zz}$, while during the AFM phase transition $V_{zz}$ jumps to a much smaller value in the low temperature AFM state. Our first-principles calculations reveal that the contribution to the EFG mainly comes from the electrons close to the Fermi level. These observations could be well understood by the magnetically induced redistribution of the charges near the Fermi level, which changes the electron distribution symmetry (hence the EFG) while has little effect on the absolute charge density at the iron nucleus during the AFM transition.

\begin{acknowledgments}
This work was supported by the National Natural Science Foundation of China under Grants No. 10975066.

\end{acknowledgments}

%\bibliography{MyRef}

\end{document}